**Empirical correlation between the interfacial Dzyaloshinskii–Moriya interaction and work function in metallic magnetic trilayers**

Yong-Keun Park,[1,2] Dae-Yun Kim,[1] Joo-Sung Kim,[1] Yune-Seok Nam,[1] Min-Ho Park,[1] Hyeok-Cheol Choi,[1] Byoung-Chul Min,[2*] and Sug-Bong Choe[1*]

[1]Department of Physics and Institute of Applied Physics, Seoul National University, Seoul, 08826, Republic of Korea.

[2]Center for Spintronics, Korea Institute of Science and Technology, Seoul, 02792, Republic of Korea.

Correspondence and requests for materials should be addressed to S.-B.C and B.-C.M (E-mail:sugbong@snu.ac.kr and min@kist.re.kr).

**The Dzyaloshinskii–Moriya interaction (DMI) generates intriguing chiral magnetic objects such as magnetic skyrmions and chiral domain walls that can be used as building blocks in emerging magnetic nanodevices. To achieve better stability and functionality of these chiral objects, it is essential to achieve a larger DMI. In this paper, we report an experimental observation that in magnetic trilayer films, the DMI strength is mainly determined by the work functions of the nonmagnetic layers interfaced with the magnetic layer. The clear correlation with the intrinsic material parameters provides a guideline for material selection to engineer the DMI strength.**

Chiral magnetic materials and the phenomena associated with them as well as the technological opportunities provided by emerging spintronic devices[1,2] have recently attracted increasing academic attention. Such chiral magnetic phenomena are caused by antisymmetric exchange interaction, which is the so-called Dzyaloshinskii–Moriya interaction (DMI)[3,4]. In magnetic thin films, a sizeable DMI generates built-in chirality of magnetic domain walls (DWs), which are essential for current-induced DW motion via spin-orbit torques (SOTs)[5,6]. The ultimate speed of the DW motion has been revealed to be also governed by the DMI strength[7]. In addition, a strong DMI can generate topological objects such as magnetic skyrmions, which can be used in high-speed and high-density digital devices such as racetrack memory and so on[1,2,8].

Because of academic interest and technological importance, numerous efforts have been devoted to understanding the DMI and achieving a larger DMI by exploring diverse combinations of materials with asymmetric-layer structures[9-13]. Although the physical origin of the DMI at the metallic interface remains unclear, identifying the intrinsic parameter that significantly affects the DMI will be of great help, which can provide a guideline in terms of both experiment and sample structure design for application as well as help in understanding the DMI.

Therefore, we are interested in the DMI at the interface between ferromagnetic metal and nonmagnetic metals in a bilayer system. The interfacial DMI in such metallic bilayers arises from a three-site indirect exchange mechanism among two atomic spins and a neighbouring atom with a large spin-orbit coupling (SOC)[2,4,14]. In antiferromagnetic crystals, the DMI is known to originate from anisotropic super-exchange interaction that linearly depends on the strength of the SOC[4]. In this case, the DMI magnitude is proportional to $(\Delta g/g)J_{\mathrm{ex}}$, where $g$ is the gyromagnetic ratio, $\Delta g$ is its deviation from the value of a free electron, and $J_{\mathrm{ex}}$ is the magnitude of the exchange interaction[4].

Whether associating this elegant picture to polycrystalline metallic systems is possible is an interesting problem. The spin-glass alloy systems shed light on the origin of the DMI in metallic systems, in which the DMI arises from the spin-orbit scattering of conducting electrons by nonmagnetic transition-metal impurities[14]. This condition inspires us to investigate the material parameters that might be correlated with *the scattering potential* at the metal–metal interface. If we can find a correlation between the DMI strength and other material parameters such as the work function, electronegativity, and SOC constant, this would provide a productive guideline to engineer the DMI strength. In this paper, we report a remarkable experimental observation that the work-function difference at the metal–metal interface is strongly correlated with the DMI strength.

**Results**

**Measurement of the SOT efficiency to quantify the DMI of trilayer structure samples**

In this study, various asymmetric trilayer structures, namely, Pt/Co/X, were prepared, where X was selected as Ti, Cu, W, Ta, Al, Ru, Pd, Au, or Pt (nine materials) to observe the relative DMI tendency of a bilayer system while keeping the bottom Pt layer the same. All the films exhibited perpendicular magnetic anisotropy.

The first step was to precisely measure the magnitude of the DMI of these Pt/Co/X samples. Figure 1(a) shows the schematic drawing of the measurement setup using a 20-μm-wide and 350-μm-long microwire sample. DMI-induced effective magnetic field $H_{\mathrm{DMI}}$ was determined by measuring SOT efficiency $\varepsilon$ with respect to in-plane magnetic field $H_x$. When current was injected, DW depinning field $H_{\mathrm{dep}}$ varied, as shown in Fig. 1(b). If the values of the depinning field were measured with respect to current density $J$, we can measure the SOT efficiency from the slope, i.e. $\varepsilon = -\partial H_{\mathrm{dep}}/\partial J$. (see Methods section for details).

Figure 2 shows the plot of $\varepsilon$ with respect to $H_x$ for the samples with different X's,

as denoted in each panel. All the plots of $\varepsilon$ exhibit three distinct regimes frequently observed in SOT-induced DW motions: two saturation regimes of the Néel-type DW configurations and a transition regime in between[15]. In the transition regime, a Bloch-type DW configuration appeared at the intercept to the abscissa (as designated by the red vertical lines) in which magnetic field $H_x^0$ exactly compensated $H_{\text{DMI}}$ (i.e. $H_x^0 + H_{\text{DMI}} = 0$). We can therefore quantify $H_{\text{DMI}}$ from these measurements. All the samples with broken inversion symmetry exhibited non-zero $H_{\text{DMI}}$, except the sample with X = Pt that showed almost zero $H_{\text{DMI}}$ owing to the symmetric-layer structure.

**Correlations between material parameters and the DMI**

Table I and II list the summary of the measured values of $H_{\text{DMI}}$, anisotropy field $H_K$, effective uniaxial magnetic anisotropy $K_{\text{eff}}$, and DMI strength $D$, as well as the literature values of work function $W$, electronegativity $\chi$, and SOC constant $\xi$, respectively. These material parameters were chosen because of their potential relationship with the electrostatic potential gradient at the interface between the Co and X layers. From the tables, we can clearly see that $H_{\text{DMI}}$ has the most distinct correlation with $W$ than the other parameters. The correlations with $W$, $\chi$, and $\xi$ are plotted and shown in Fig. 3, which clearly shows that the clearest correlation exists in Fig. 3(a) than in the other figures. This observation suggests that the work function plays a significant role in the generation of $H_{\text{DMI}}$.

From the concept of potential gradient at the interface, the electronegativity may also be presumed to have a relationship trend with the DMI, as shown in Fig. 3(b), with a rough correlation between $\chi$ and $H_{\text{DMI}}$. Although the electronegativity difference between Co and X also implies a potential gradient at the interface, it is rather relevant to atomic and/or molecular systems but less relevant to metallic bilayer systems. In addition, because the spin-orbit-scattering-mediated spin-chiral effect may plays a leading role in our trilayer system, the relationship tendency between the DMI and work function is more obvious instead of the case of electronegativity.

Another very important factor in determining the strength of the DMI is the SOC. A simple conjecture is that strength $D$ of the DMI could be linearly proportional to the strength of the SOC similar to that in antiferromagnetic crystals[4]. However, the experimental correlation between $H_{DMI}$ and $\xi$ is found to be pretty scattered, as shown in Fig.3(c). Because the present values of $\xi$ corresponds to the atomic SOCs from the literature, if more specific and precise SOC values relevant to the metallic bilayer systems were available, we may possibly see a more accurate relationship between the DMI and SOC.

The magnitude of $D$ is estimated from the relationship $D = \mu_0 H_{DMI} M_S \lambda$ with Bloch-type DW width $\lambda$ $(= \sqrt{A/K_{eff}})$[7,16], where $A$ is the exchange stiffness. The Co value of $A$ ($2.2 \times 10^{-11}$ J/m) and $M_S$ of Co ($1.4 \times 10^6$ A/m) are used in the estimation for qualitative comparison[16,17]. The effective anisotropy is quantified from the relationship $K_{eff} = M_S H_K / 2$. We presume that the scattering potential is associated with the work function difference $\Delta W$ atthe Co/X interface. Figure 4 shows a summary of the $D$ values as a function of $\Delta W$ ($\equiv W_X - W_{Co}$). A correlation is clearly observed between $D$ and $\Delta W$. The electron redistribution across the junction is well known to generate an electrostatic potential gradient to balance the Fermi level between the materials at the metallic heterostructure. The work function difference at the Co/X interface leads to electrostatic potential gradient, which in turn gives rise to a sizable spin-orbit scattering of conduction electrons at the interface[14].

**Discussion**

The signs of the DMI are all negative in the samples of the Pt/Co/X trilayer structure except for X=Pt. In the case of X=Pt, the sign is not obviously confirmed, but its magnitude is very small. In a symmetric structure, the sign and magnitude of the DMI can be determined according to the state of each interface. In addition, in a previous experimental result, the sign of the DMI in Pt/Co/Pt has been reported as positive[18]. It is known that a strong SOC exists at the Co/Pt interface, and Ref. [12] predicted that the DMI strength is very large at the Co/Pt bilayer interface. From this result, one may infer that the sign of the DMI may be determined in our trilayer samples by the underlayer Pt. Because our measurement is the sum of the effects on the DMI of the two interfaces, the effect of each interface cannot be independently observed. Even if the underlayer Pt is dominant on the total DMI, directly knowing the sign and magnitude of the DMI at Co/X bilayer interface is difficult. However, the relative strength of the DMI between the Pt/Co/X trilayers remains significant.

A potential gradient is developed when a metal−metal interface is formed. However, together with it, the charge redistribution and change in the Fermi level position, i.e. chemical potential, can also occur at the interface. From the result of the electronegativity with the DMI, the potential gradient may still be a dominant factor in the tendency. However, several other effects may also be associated with the work function. Therefore, in a real situation, a complicated mechanism appears to be present in the system, and the exact magnitude of $D$ could be associated with the sophisticated band structures. Still, the remarkable point is that a rough but general trend exists in the determination of the DMI strengths in connection with the work function in a Pt/Co/X trilayer structure.

In summary, we have presented an experimental observation that the DMI strength is strongly correlated with the work function. Such strong correlation might be related to the spin-orbit scattering in the electric potential gradient caused by the work function difference at the interfaces. The present correlation suggests that the DMI strength can be engineered via material selection following the guidelines related to the intrinsic material parameters.

**Methods**

**Sample fabrication and detail structure**

The detailed film structure is 5-nm Ta/2.5-nm Pt/0.9-nm Co/2.5-nm X/1.5-nm Pt, which is deposited by dc-magnetron sputtering on Si wafers with a 300-nm $SiO_2$ layer. The Co layer thickness is kept constant for direct comparison among the samples. The lowermost Ta layer is a seed layer used to enhance the crystallinity of the films, and the uppermost Pt layer is a protective layer used to prevent oxidation. The samples are patterned to 20-μm-wide and 350-μm-long microwires by photolithography and ion-milling process, as shown in Fig. 1(a).

**Measurement of the SOT efficiency**

SOT efficiency $\varepsilon$ is measured from the depinning field of the DWs. The measurement procedure is described as follows. A DW is created at a position inside the microwire adjacent to the DW writing electrode [white vertical line in Fig. 1(a)]. Then, an out-of-plane magnetic field is swept until the DW moves from the initial position. By repeating this procedure using various current biases, as shown in Fig. 1(b), we can measure depinning field $H_{\text{dep}}$ as a function of current density $J$ for a certain value of $H_{\text{x}}$. From the linear dependence of $H_{\text{dep}}$ on $J$, as shown in Fig. 1(c), we can quantify $\varepsilon$ using the relationship $\varepsilon = -\partial H_{\text{dep}}/\partial J$. The measurement is repeated for different $H_{\text{x}}$ values. Then, the $\varepsilon$ data with respect to $H_\text{x}$ are obtained, as shown in Fig. 2.

**Acknowledgements**

This work was supported by grants from the National Research Foundations of Korea (NRF) funded by the Ministry of Science, ICT and Future Planning of Korea (MSIP) (2015R1A2A1A05001698 and 2015M3D1A1070465). Y.-K.P. and B.-C.M. were supported by the National Research Council of Science & Technology (NST) (Grant No. CAP-16-01-KIST) by the Korea government (MSIP).

**Author contributions**

Y.-K.P. planned and designed the experiment, and S.-B.C. supervised the study. Y.-K.P., M.-H.P., H.-C.C., and B.-C.M. prepared the samples. Y.-K.P., D.-Y.K., J.-S.K., and Y.-S.N. carried out the measurement. S.-B.C., B.-C.M., and Y.-K.P. performed the analysis and wrote the manuscript. All authors discussed the results and commented on the manuscript.

**Additional information**

Correspondence and requests for materials should be addressed to S.-B.C. and B.-C.M.

**Competing financial interests**

The authors declare no competing financial interests.

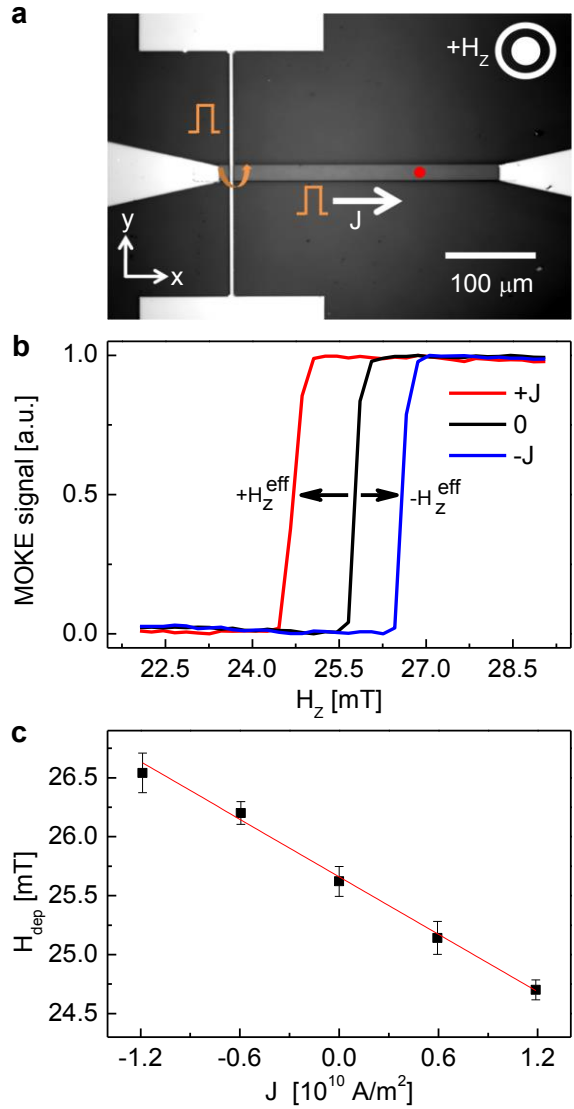

**Figure 1 | Measurement of SOT efficiency.** (**a**) Schematic drawing of the measurement setup with the optical image of a sample. The grey horizontal rectangle is the 20-μm-wide and 350-μm-long microwire, and the white areas are the 5- and 100-nm-thick Ti/Au electrodes. The white vertical line shows the electrode for the DW writing. The red circular dot indicates the position of the laser spot for the magneto-optical Kerr effect (MOKE) signal detection. The polarities of the external magnetic field and current are designated inside the figure. (**b**) Normalised MOKE signals with respect to $H_z$ for different current biases $J =$ (red) $+1.2 \times 10^{10}$, (black) 0, and (blue) $-1.2 \times 10^{10}$ A/m$^2$. (**c**) $H_{\text{dep}}$ with respect to $J$ for a

fixed in-plane magnetic field bias (-300 mT). The red line shows the best linear fitting.

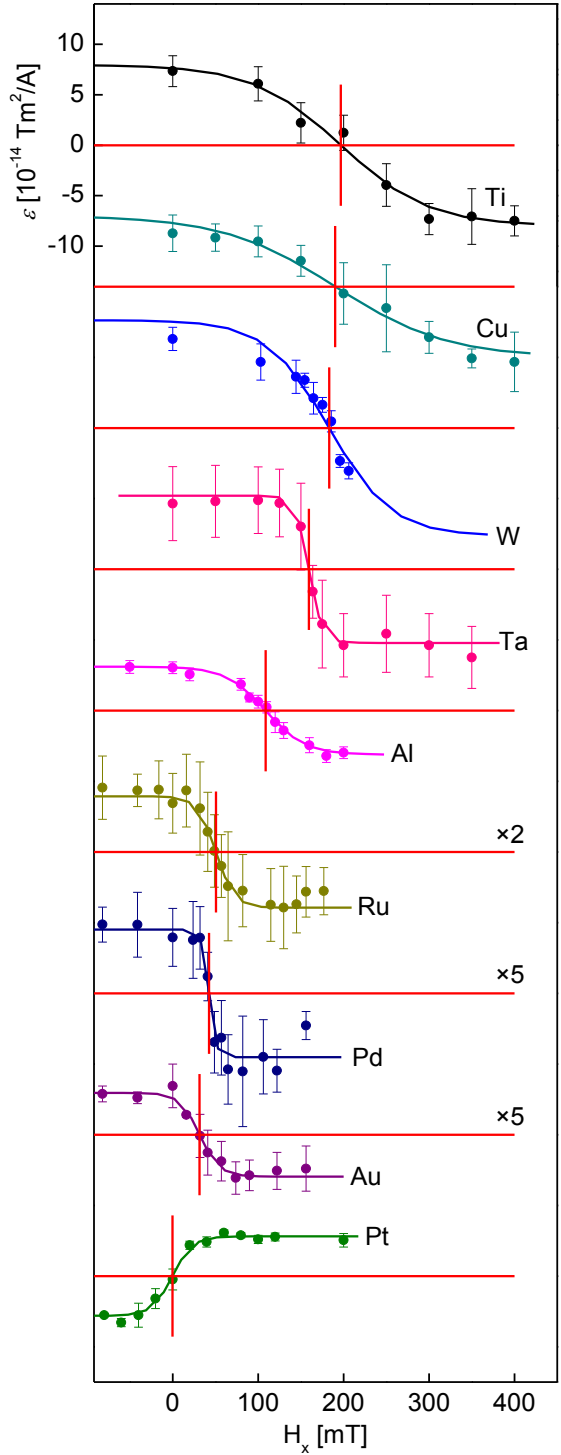

**Figure 2 | SOT efficiency of Pt/Co/X.** Plots of $\varepsilon$ with respect to $H_x$ for different X's as denoted in the figures. The curved solid lines are the best fitting for eye guide. The red horizontal lines show the axis of $\varepsilon = 0$ for each measurement. The red vertical lines indicate

the positions of $H_x^0$ for $\varepsilon$ is zero. The values of $\varepsilon$ for X=Ru, Pd, and Au are magnified for better resolution of the data with the amplification factors denoted in the figure.

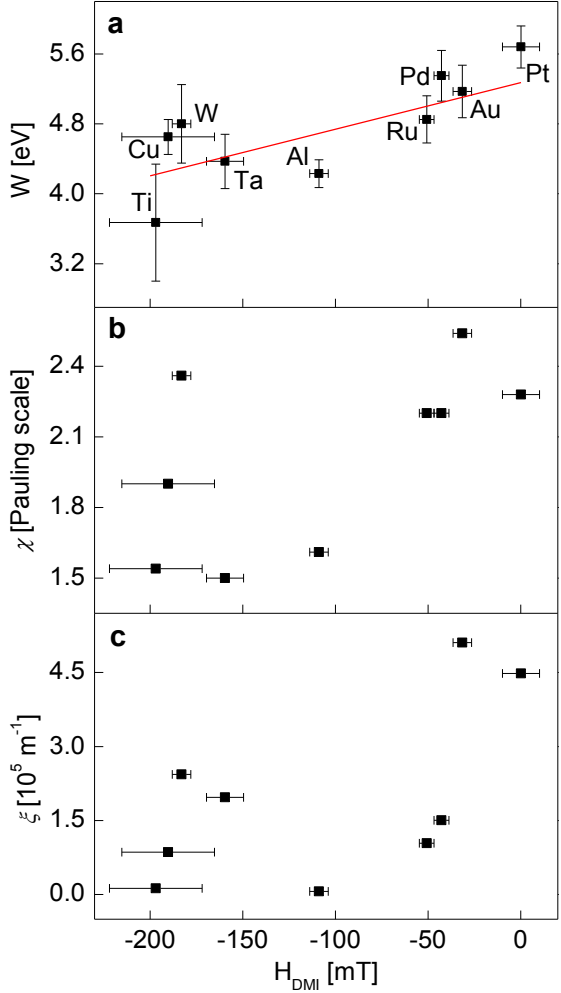

**Figure 3 | Parameter relationships with $H_{\text{DMI}}$.** (**a**) Work function $W$, (**b**) electronegativity $\chi$, and (**c**) SOC constant $\xi$ with respect to $H_{\text{DMI}}$ for Pt/Co/X samples with different X's, as denoted in the figure. The red line in (**a**) shows the best linear fitting for eye guide. In each plot, $H_{\text{DMI}}$ is determined using the $\varepsilon$ measurement from the data shown in Fig. 2, and $W$, $\chi$, and $\xi$ are the literature values from Refs. [19]–[45]. The error bars of $H_{\text{DMI}}$ corresponds to the experimental accuracy for several repeated measurements. The error bars of $W$ are obtained as the standard deviation of several different values from the references.

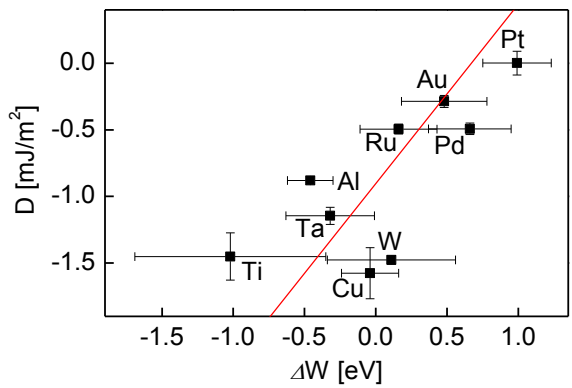

**Figure 4 | Correlation between *D* and *ΔW*.** Plot of *D* with respect to *ΔW* for Pt/Co/X samples with different X's, as denoted near each data symbol. The red line shows the best linear fitting. The error bars of *D* are obtained from the statistical summation of the experimental accuracy of $H_{\text{DMI}}$ and $H_{\text{K}}$ determination. The error bars of *W* are obtained as the standard deviation of several different values from the references.

**Table I | $H_{DMI}$, $H_K$, $K_{eff}$, and $D$ for Pt/Co/X samples with material X as well as the quantified values from measurement.**

| X | $H_{DMI}$ [mT] | $H_K$ [T] | $K_{eff}$ [$10^5$ J/m$^3$] | $D$ [mJ/m$^2$] |
|---|---|---|---|---|
| Ti | -197±25 | 1.13±0.013 | 7.94±0.09 | -1.45±0.18 |
| Cu | -190±25 | 0.90±0.021 | 6.28±0.15 | -1.58±0.19 |
| W | -183±5 | 0.95±0.012 | 6.62±0.08 | -1.48±0.03 |
| Ta | -160±10 | 1.20±0.018 | 8.37±0.13 | -1.15±0.06 |
| Al | -109±5 | 0.94±0.049 | 6.61±0.35 | -0.88±0.02 |
| Ru | -51±4 | 0.65±0.006 | 4.54±0.04 | -0.49±0.04 |
| Pd | -43±4 | 0.46±0.006 | 3.24±0.04 | -0.49±0.04 |
| Au | -32±5 | 0.75±0.006 | 5.22±0.04 | -0.29±0.04 |
| Pt | 0±10 | 0.79±0.010 | 5.51±0.07 | 0.00±0.09 |

**Table II | $W$, $\chi$ (Pauling scale), and $\xi$ of material X.** The values of $W$ are given as the mean values with the standard variation from Refs. [22]–[43]. The Co value of $W$ is 4.69±0.27 eV [19-22]. Superscripts [*] and [**] indicate relevant Refs. [44] and [45], respectively.

| X | $W$ [eV] | $\chi$[*] [Pauling] | $\xi$[**] [$10^4$ m$^{-1}$] |
|---|---|---|---|
| Ti | 3.67±0.67 | 1.5 | 1.2 |
| Cu | 4.65±0.20 | 1.9 | 8.6 |
| W | 4.8±0.56 | 2.4 | 24.3 |
| Ta | 4.37±0.31 | 1.5 | 19.7 |
| Al | 4.23±0.16 | 1.6 | 0.6 |
| Ru | 4.85±0.27 | 2.2 | 10.4 |
| Pd | 5.35±0.29 | 2.2 | 15.0 |
| Au | 5.17±0.30 | 2.5 | 51.0 |
| Pt | 5.68±0.24 | 2.5 | 44.8 |